# Spectral compression of single-photon wave packets by sum-frequency conversion in slow-light waveguides

Michael G. Raymer

*Department of Physics and Center for Optical, Molecular and Quantum Science,*
*University of Oregon, Eugene, Oregon 97403, USA*
raymer@uoregon.edu

**Abstract:** A slow-light scheme is proposed for simultaneous frequency conversion and spectral compression of a weak optical pulse, which may be in any quantum state including a single-photon state. Such a process plays crucial roles in a number of schemes for constructing quantum networks. Assuming that appropriate slow-light waveguides can be fabricated, theoretical modeling shows that a 3-ps pulse can be converted by sum-frequency generation into a pulse with duration in the ns regime with a corresponding spectral compression factor of the order of 1000 and a useful intrinsic efficiency up to 83%. Independent of the input pulse shape, the converted pulse will have a near-exponential rising shape, which is suitable for temporal-mode matching into an optical cavity.

## 1. Introduction

Time-frequency quantum-state engineering of single photons and biphotons is a key tool in quantum information science. [bre15, mat16, don16b, kar17, ave17, ans18, ray20, fab20] The temporal duration and spectral bandwidth of photons produced by various sources (e.g. spontaneous parametric down conversion, spontaneous four-wave mixing, or intracavity single atoms or quantum dots) are often not well matched in frequency, spectral width and wave-packet shape for interacting with a targeted quantum memory. Therefore, techniques for temporal-spectral transformations are needed for advancing the state of the art.

For example, a recently proposed scheme ('ZALM') uses spectral multiplexing of biphotons generated by spontaneous parametric down conversion to increase the rate of entanglement distribution between remote quantum memories consisting of cavity-based color centers in diamond [che23]. The challenge for this scheme is that passive spectral demultiplexing of the wide-band biphotons results typically in photon wave packets with bandwidth 50 to 100 GHz, while the cavity-color-center memories require input photon bandwidth at least 100 times smaller. [ray24, sha24] The ZALM scheme requires a larger spectral compression of photon wave packets than has been demonstrated to date.

Quantum frequency conversion by sum-frequency generation (SFG) [kum90] and four-wave mixing [mcg10] are proven methods for, not only shifting the carrier frequency of an optical wave packet (a single-photon packet or a coherent-state pulse) to a targeted value, but also for implementing temporal-spectral transformations including spectral compression. [mck12, lav13, don16a, all17] The challenge has been to carry out the needed transformations with efficiency approaching unity and with negligible background noise added. There are also electrooptic methods for spectral compression without large frequency conversion, which we call intra-band conversion and won't consider in this study. [kar17]

Spectral compression accompanied by inter-band frequency conversion has been achieved to date by two schemes. Pre-chirping the signal photon and pump with opposite signs followed by SFG in a thin second-order-nonlinear-optical crystal has been shown to compress the bandwidth of a single-photon packet by a factor of 40 with a conversion efficiency less than 1%, with room for moderate improvements in efficiency. [lav13] While higher compression factors could be achieved using larger chirps, the required pump power scales unfavorably with the amount of chirp introduced because the pump's peak power decreases accordingly.

Another scheme uses unchirped pulses and a longer crystal engineered for group-velocity matching, as explained below, to achieve a compression factor of 7.5 with efficiency around 20%. [all17] The present proposal extends this idea by introducing slow-light techniques.

In this paper a scheme is introduced for high-efficiency spectral compression of optical wave packets based on SFG in a second-order nonlinear-optical crystal waveguide using slow-light techniques. The waveguide dispersion is engineered for group-velocity matching of the input photon wave packet and the pump laser pulse and engineered for slow light of the SFG pulse. Group-velocity matching allows the use of crystals that are long enough to provide high conversion efficiency using moderate pump power while maintaining a wide acceptance bandwidth, as shown in [all17]. The innovation here is the recognition that engineering a much slower group velocity for the SFG enables even greater spectral compression and higher efficiency and at the same time creates a single-photon SFG pulse shape that is suitable for temporal-mode matching to a cavity-based quantum memory. This novel arrangement can ideally reach a spectral compression factor approaching 1000 with conversion efficiency greater than 80%. Through analytical and numerical solutions of the three-wave mixing equations we gain insight into the dynamics leading to such benefits.

## 2. Single-sideband group-velocity-matched SFG

SFG is the process by which a weak 'signal' field having canter angular frequency $\omega_s$ and (one-dimensional) spatial-temporal amplitude $A_s(z,t)$ mixes nonlinearly in a $\chi^{(2)}$ medium with a strong 'pump' field having frequency $\omega_p$ and amplitude $A_p(z,t)$ to generate a sum-frequency 'register' field having frequency $\omega_r = \omega_s + \omega_p$ and amplitude $A_r(z,t)$. (The name 'register' arises for the SFG when the process is used to implement quantum-logic operations. [bre15])

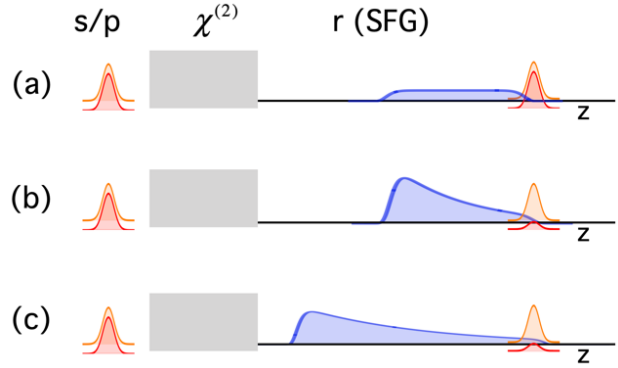

Fig. 1. Pump *p* and input signal *s* travel with same group velocity while SFG *r* travels more slowly as a result of wave-guide dispersion. (a) low conversion efficiency, (b) high conversion efficiency, (c) same as (b) but with the SFG pulse slowed further using slow-light techniques. The input pulses are shown as unit-normalized, so their heights do not represent their energy content. Note that the horizontal axis is position *z* for fixed a time.

The SFG process with group-velocity matching and slow light is illustrated in **Fig. 1**. The pump *p* and input signal *s* travel with the same group velocity while the SFG *r* travels more slowly. In [red13, red17] this scenario is called 'single-sideband group-velocity-matching' (SSGVM). If the conversion efficiency is very low, the SFG is created as a stretched pulse with a quasi-rectangular shape, as in Fig. 1(a). As the conversion efficiency is increased by increasing the pump power, the leading edge of the pulse (corresponding to the latest generated portion) becomes suppressed as a result of depletion of the input signal pulse, as in Fig. 1(b). The resulting pulse shape is similar to a rising exponential, which is known to be optimal for loading into a cavity. [sto09]

While the behaviors in Figs. 1(a) and 1(b) are known from previous works [red13, all17], to date it has not been recognized that extreme slowing of the SFG pulse can lead to important performance improvements in spectral compression and pulse reshaping. The prior work on spectral compression relied on the SFG slowness achievable through the bulk material properties and engineering the transverse dimensions of the waveguide. [bre14, all17]

For very short pump and signal pulses, the length of the SFG pulse is given by the difference of the times-of-flight through the medium, $(\beta_r - \beta_s)L$, where $\beta_j$ is the inverse of the group velocity (called group slowness) of pulse *j*. [mck12] Part (c) of Fig. 1 illustrates the SFG pulse lengthening when the SFG pulse is slowed more than in Part (b). For example, if $\beta_p = \beta_s = (c/2.5)^{-1}$, $\beta_r = (c/10)^{-1}$, (where *c* is the vacuum speed of light) and $L = 40$ *mm*, and the input pulse duration is of the order $10$ *ps*, then the generated SFG pulse has duration $1$ *ns*. The SFG pulse is a coherent wave packet so its spectral bandwidth is given by the usual Fourier transform relationships, meaning that the photon's bandwidth is compressed by a factor approaching 1000.

Here I propose to control the pulses' group velocities using the techniques of 'slow light,' a phenomenon that is well-studied in many linear-optical and nonlinear-optical contexts. For review, see [bab08, sch10, boy11, sko16]. In optical fibers or waveguides, group velocities can be engineered by introducing a linear-refractive-index Bragg grating. Two general approaches have been used: a photonic bandgap can be introduced uniformly throughout the waveguide's length with the band edge tuned near the optical frequency targeted for slowing. The bandgap can be engineered using either gratings superimposed on a waveguide structure or as an integral part of the design of a 2D photonic-crystal waveguide. The other approach is similar but introduces a sequence of near-resonant Bragg cavities along the length of the waveguide, a CROW (coherent resonant optical waveguide) structure. In the present application the goal would be to slow the SFG pulse considerably (slowing factor > 10) while leaving the pump and signal pulses little affected. Further comments on the feasibility of achieving the needed slow group velocities for this application are postponed to a later section.

It is also worth noting that the SSGVM scenario is precisely what is needed for effective operation of the 'quantum pulse gate,' a technique for selecting a single (targeted) temporal mode from a pulse consisting of several temporal modes. [eck11, red14, bre15, man16, red17, red18] In a follow-up paper we will explore how slow-light techniques can improve the operation of the quantum pulse gate.

In some applications that use polarization qubits it is required to frequency convert and spectrally compress a signal that contains an arbitrary and unknown state of polarization (H, V, elliptical, etc.). We note that frequency conversion of arbitrary polarization states has been demonstrated using a Sagnac interferometer containing a single second-order nonlinear crystal that converts only one targeted polarization. [rik18, yan19] The scheme works by splitting the signal into two fields and using polarization-controlling elements inside the interferometer to recombine the fields after frequency conversion. Analogous methods use a Mach-Zehnder interferometer. [li19] Here we assume that such techniques would be employed if needed and so we model SFG for a single polarization.

It should be noted that spectral compression has been predicted and observed in second harmonic generation, in which fs pulse were stretched to ps, with the corresponding spectral compression. [mar07] In that case, group-velocity matching is automatic because the pump and input signal are one and the same. In contrast, here we consider conversion from ps pulse to ns pulses using SFG.

### 3. Model set up for SFG with engineered group velocities

To model the SFG process in the case of a second-order nonlinear-optical medium, we define scaled electric-field amplitudes as $E_j(z,t) = A_j(z,t) exp(-i\omega_j t + ik_j z)$ for $j = p, s, r$. The slowly varying amplitudes $A_{r,s}(z,t)$ ($s^{1/2}$) are called 'temporal modes' and obey the same Maxwell-equation dynamics as do classical field amplitudes. [ray23] The transformation between signal and SFG fields (that is, modes) is taken to be unitary (energy conserving and reversible). That is, we neglect dissipative losses and depletion of the classical-like pump field. Thus, the transformation is equivalent to a lossless beam-splitter transformation between temporal modes, a fact that leads to several interesting applications. [red18, cle16, ans18] (The beam-splitter analogy allows one to think of the r pulse as the 'reflected' pulse.) The temporal modes can be viewed as receptacles for any quantum state of light (single-photon, squeezed, coherent-state) as long as the state contains far fewer photons than does the pump pulse, so pump depletion can be ignored.

For an interaction that is phase-matched for the three central frequencies $\omega_{s,p,r}$, and assuming the pump is undepleted, the equations of motion for the amplitudes, in the slowly-varying-envelope approximation, are [red14]

$$\begin{aligned}(\partial_z+\beta_r\partial_t+i\beta_{2r}\partial_t^2)A_r(z,t) &= i\gamma A_p(z,t)A_s(z,t) \\ (\partial_z+\beta_s\partial_t+i\beta_{2s}\partial_t^2)A_s(z,t) &= i\gamma A_p^*(z,t)A_r(z,t) \\ (\partial_z+\beta_p\partial_t+i\beta_{2p}\partial_t^2)A_p(z,t) &= 0\end{aligned} \quad (1)$$

where $\beta_j = \partial_\omega k(\omega)_{\omega_j}$ is the inverse of the group velocity ($\text{s/m}$) and $\beta_{2j} = (1/2)\partial_\omega^2 k(\omega)_{\omega_j}$ is (one-half) the group-delay dispersion (GDD) per unit length ($\text{s}^2/\text{m}$) for the $j$ pulse. $\gamma$ is the mode-coupling strength ($\text{s}^{1/2}/\text{m}$), proportional to the effective $\chi^{(2)}$ nonlinearity of the medium and to the square root of the pump pulse energy (we assume the input pump and signal amplitude functions are square-normalized). We have assumed that the three fields have polarizations fixed for optimal phase matching, and hence treat them as scalar functions.

To appreciate the significance of the coefficients in Eq.(1), write the well-known form of the phase mismatch for three-wave mixing,

$$\Delta k(\omega,\tilde\omega) = k(\omega) - k(\tilde\omega) - k(\omega-\tilde\omega) - \frac{2\pi}{\Lambda}, \quad (2)$$

where $k(\omega)$ is the dispersion relation for the medium, $\omega,\tilde\omega$ are the (angular) frequencies of the $r$ and $s$ fields, respectively, with the pump frequency constrained to equal $\omega-\tilde\omega$, and $\Lambda$ is the poling period ($\text{m}$) for quasi-phase matching that enables phase matching for the central frequencies, $\Delta k(\omega_r,\omega_s)=0$. Using a truncated Taylor series, $k(\omega) \approx k_j + \beta_j(\omega-\omega_j) + \beta_{2j}(\omega-\omega_j)^2$ for each spectral band, the phase mismatch becomes

$$\begin{aligned}\Delta k(\omega,\tilde\omega) &= (\omega-\omega_r)(\beta_r-\beta_p) - (\tilde\omega-\omega_s)(\beta_s-\beta_p) \\ &+ \beta_{2r}(\omega-\omega_r)^2 - \beta_{2s}(\tilde\omega-\omega_s)^2 - \beta_{2p}(\omega-\tilde\omega-\omega_r+\omega_s)^2.\end{aligned} \quad (3)$$

If the quadratic contributions are negligible, then we see that perfect phase matching, $\Delta k(\omega,\tilde\omega)=0$, dictates a line in the $\omega,\tilde\omega$ plane given by $(\omega-\omega_r) = (\tilde\omega-\omega_s)(\beta_s-\beta_p)/(\beta_r-\beta_p)$ with slope $(\beta_s-\beta_p)/(\beta_r-\beta_p)$. Thus, if the waveguide is dispersion engineered [xin22, pol24] such that the signal and pump are group-velocity matched, $\beta_s = \beta_p$, the slope is zero and the phase-matching line is horizontal, as pointed out in the SFG context in [all17].

In the case that $\beta_s = \beta_p$ and the quadratic contributions are negligible, the spectral width of the phase-matching function with respect to $(\omega - \omega_r)$ is estimated from Eq.(3) as $\delta_r = \pi / (\beta_r - \beta_p)L$, which (as we will see below) is inversely proportional to the temporal duration of the converted pulse. Thus, the greater the velocity mismatch of pump and SFG pulses, the narrower the spectrum will be.

In **Fig.2** we plot the phase-matching function $sinc[\Delta k(\omega, \tilde{\omega})L/2]$ in periodically-poled lithium niobite (PPNL) for a signal with wavelength 1560 nm and pump wavelength 907 nm, giving SFG wavelength 574 nm, which is matched to the charge-neutral nitrogen vacancy center (NV0) in diamond. [poem] Using the known temperature-dependent Sellmeier dispersion equation for lithium niobate at 300 C [jun97], we show three cases, all Type-0 with quasi-phase matching of the zero-order propagation constants, $k_r - k_s - k_p - 2\pi/\Lambda = 0$, with $\Lambda = 8.4\,\mu m$. (a) bulk PPLN with $L = 1.5$ mm, (b) bulk PPLN with $L = 27$ mm, and (c) dispersion-engineered PPLN approximated by using the same parameters as bulk PPLN except assuming SSGVM, that is replacing the value of $\beta_s$ by the value of $\beta_p$. SFG is generated within the region (outlined in pink) of mutual overlap of the input signal spectrum (vertical dark dashed lines), the pump spectrum (light dashed lines along the diagonal to account for energy conservation) and the phase-matching function (yellow-to-white being optimal phase matching).

Fig.2a shows that a thin PPLN crystal achieves a wide input acceptance bandwidth and the output SFG bandwidth is also wide, more or less independent of the pump bandwidth. This case is similar to that in [lav13] where a chirped pump (not shown) can lead to spectral compression in the SFG, but because the crystal is thin the conversion efficiency is limited for moderate pump powers. Fig.2b shows that a longer bulk crystal (which might be useful for achieving higher efficiency) has a narrow input acceptance bandwidth making it unable to accept a wide-band signal for spectral compression. Fig.2c shows that a longer PPLN crystal with waveguide-engineered group-velocity matching gives a wide acceptance bandwidth and a much narrower SFG bandwidth, thus providing moderate spectral compression with no need to chirp the pulses, as demonstrated in [all17].

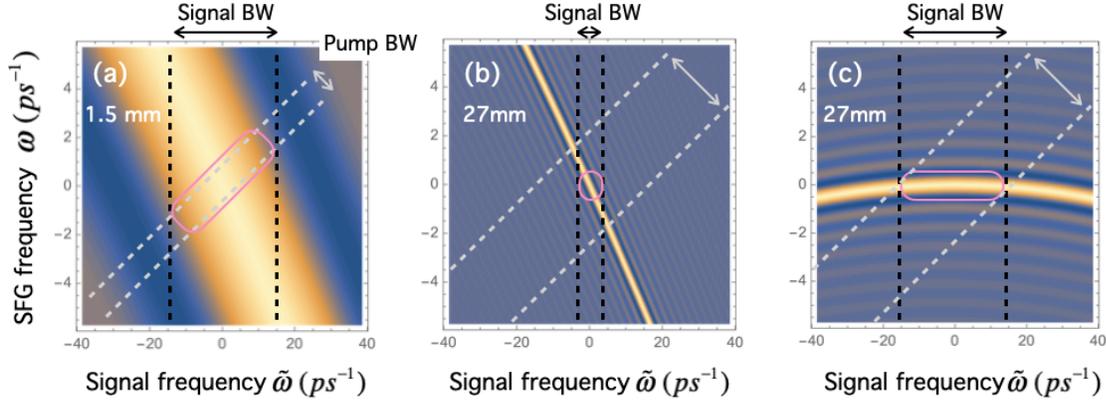

Fig. 2. Phase-matching function for the process 1560 nm + 907 nm $\to$ 574 nm. Light yellow indicates optimal phase matching. (a) $L$ = 1.5 mm bulk PPLN, (b) $L$ = 27 mm bulk PPLN, (c) $L$ = 27 mm waveguide-engineered PPLN with the same parameters as bulk except assuming $\beta_s = \beta_p$.

The input signal and pump spectra are indicated as regions demarcated by dashed lines. Units of angular frequency are 1/ps. In terms of wavelengths, the horizontal axis runs from 1514 nm to 1614 nm and the vertical from 573 nm to 575 nm, as in [all17].

## 4. Analytical solutions for SSGVM frequency conversion

We consider the case that the pulses are long (narrow-band) enough that for a given medium length any higher-order group-delay dispersion (the curvature in Fig. 2b) is insignificant. The criterion for such to be the case is seen by taking the Fourier transform of the left-hand sides of Eq.(1), giving $(\partial_z - i\omega\beta_j - i\omega^2\beta_{2j})\tilde{A}_j(z,\omega)$. Because $\omega$ is the frequency measured from the carrier frequency and the extent of the pulse's frequency width is roughly the inverse of the pulse duration $T$ (assuming no pulse chirp), the GDD term can be neglected if $\beta_{2j}L/T^2 << 1$. We have verified by numerical solutions of the full Eq.(1) that neglecting the higher-order terms is valid for the cases presented in this paper.

Assuming that the pump field is strong and essentially unaffected by the nonlinear interaction, we can replace $A_p(z,t) \to A_p(t - \beta_p z)$.

For analytical solutions, we restrict ourselves to the SSGVM case. The result, valid for arbitrary SFG conversion efficiency, is [mck12, red13]

$$A_r(L,t) = i\bar{\gamma} \int_{-\infty}^{\infty} dt' \, \Pi(t-t') J_0\left(2\bar{\gamma}\sqrt{\zeta(t,t')(\beta_r L - t + t')}\right) |A_p(t')| e^{i\theta(t')} |A_s(0,t')| e^{i\phi(t')}. \quad (4)$$

where the temporal windowing function, which is determined by the crystal's length and the group velocity mismatch, is

$$\Pi(t-t') = \begin{cases} 1 & \text{for } \beta_s L < (t-t') < \beta_r L \\ 0 & \text{otherwise} \end{cases}, \quad (5)$$

which defines a window of duration $(\beta_r - \beta_s)L$, assumed to be positive. $J_0(x)$ is the Bessel function of zero order and the integrated pump power is defined as

$$\zeta(t,t') = \int_{t'}^{t-\beta_s L} |A_p(x)|^2 dx \;. \tag{6}$$

We expressed the pump and input signal fields in terms of positive amplitudes and time-dependent phases as $|A_p(t')|e^{i\theta(t')}$ and $|A_s(0,t')|e^{i\phi(t')}$. And we defined a scaled gain coefficient, $\bar{\gamma} = \gamma/(\beta_r - \beta_s)$, where $\gamma$ has units $s^{1/2}/m$. This solution generalizes that in [don16a] which is valid only for crystals that are thin enough (small $L$) that group velocity mismatch can be neglected.

We see from Eq.(4) that the instantaneous pump temporal phase adds directly to the temporal phase of the input signal. [red13] So, if the pump chirp is opposite to the signal chirp, that is $\theta(t) = -\phi(t)$, the signal chirp is removed, resulting in spectral compression, as in [lav13, don16a]. And we note that in this case the dynamics and conversion efficiency are the same as if the signal and pump were not chirped, with the caveat that for very long pump pulses the conversion efficiency tends to be small.

As a special case, the solution Eq.(4) can be analyzed in the limit of low conversion efficiency. Setting $J_0(x) \approx 1$, the solution in this case is a convolution of the product of the input and pump pulses with the windowing function,

$$A_r(L,t) = i\bar{\gamma} \int_{-\infty}^{\infty} dt' \, \Pi(t-t') e^{i[\theta(t')+\phi(t')]} |A_p(t')| |A_s(0,t')| \tag{7}$$

Assuming that the signal-pulse chirp is perfectly cancelled by the pump anti-chirp, $\theta(t) = -\phi(t)$, (or that neither had any chirp to begin with), the spectrum of the generated SFG pulse is determined by two factors: the temporal durations of the chirped signal and pump pulses, and the time-windowing effect of the interaction in the medium. In the short-pulse, thick-medium limit with large group-velocity mismatch between SFG and pump pulses, the generated SFG pulse has a nearly rectangular shape, as in Fig.1(a), with duration equal to the duration of the windowing function, which equals $(\beta_r - \beta_s)L$, that is the difference of the times of flight of the signal (and the pump) field and the SFG field. In the long-pulse, thin medium limit, the generated SFG pulse has duration slightly less than the pump and input pulses (assumed to be identical).

Of primary interest here is the limit in which the SFG travels much slower than the pump and signal, which travel together, that is $\beta_r \gg \beta_s = \beta_p$. If the input pulses are coincident, have the same shape, have no chirps (or have opposite chirps), and have durations much shorter than all other time scales (in particular the relative delay time $(\beta_r - \beta_s)L$), Eq.(4) can be well approximated as

$$A_r(L,t) \approx i\bar{\gamma}\, \Pi(t-T_{center})\, J_0\left(\bar{\gamma}\sqrt{2(\beta_r L + T_{center} - t)}\right), \tag{8}$$

where $T_{center}$ is the central time of the signal and pump pulses at the input of the medium. In the limit of moderate gain and a long medium, this solution is consistent with the form illustrated in Fig.1(b, c), with a rising amplitude in time (at fixed $z$ position). As we will see, this is the regime where slow-light techniques could give ideal results for spectral compression with high efficiency and useful pulse shape.

It's important to note that the Bessel function in Eq.(8) begins to oscillate when its argument exceeds the value of it first zero, approximately 2.40483. Using $t = T_{center} + \beta_s L$, we can place the Bessel zero at the start of the SFG pulse's time window (left-most point of the rectangular time window), $t = T_{center} + \beta_s L$. Then, using $\bar{\gamma} = \gamma / (\beta_r - \beta_s)$, the corresponding "critical value" of the coupling parameter, above which oscillations appear, is

$$\gamma_{crit} = \sqrt{2.89(\beta_r - \beta_s)/L}\ . \tag{9}$$

Another special case is when all three pulses have equal or nearly equal group velocities (as with a sufficiently thin medium and/or very long pulses). Then we can set $\beta_r = \beta_s = \beta_p = \beta$ and solve Eq.(4) to find (See **Appendix A**),

$$\begin{aligned}A_r(L,\tau) &= ie^{i[\theta(\tau)+\phi(\tau)]}\left|A_s(0,\tau)\right|\sin(\gamma\,|A_p(\tau)|\,L)\\ A_s(L,\tau) &= e^{i\phi(\tau)}\left|A_s(0,\tau)\right|\cos(\gamma\,|A_p(\tau)|\,L)\ ,\end{aligned} \tag{10}$$

where the retarded time variable is $\tau = t - \beta L$. This result is consistent with discussions in [red13] and [don16a]. (This result can also be derived by taking the limit $\beta_r \to \beta_s = \beta_p = \beta$ in Eq.(4). [red13]) Again, that if the pump pre-chirp is opposite to the signal pre-chirp, $\theta(t) = -\phi(t)$, the signal chirp will be removed, resulting in spectral compression. We see that the signal and SFG exchange energy in an oscillatory manner in each time slice (fixed value of $\tau$) but the rate of exchange is different for different values of $\tau$, leading to complicated space-time oscillations that prevent full exchange of energy and restricting the amount of spectral compression that can be acheived. This effect can be minimized by making the pump pulse duration much greater than the signal pulse at the expense of requiring a larger total pump pulse energy.

We assume a Gaussian pump pulse for which we can write the analytical solution for its pulse evolution from Eq.(4), allowing for dispersion (but no chirp),

$$A_p(z,t) = \frac{\sqrt{T_p}}{i(2\pi)^{1/4}\sqrt{T_p^2 + i\beta_{2p} z}}\exp\left[-\frac{i\left(t - T_{center} - z\beta_p\right)^2}{4\left(iT_p^2 + \beta_{2p} z\right)}\right] \tag{11}$$

where $T_{center}$ is the time of peak amplitude at the input $z=0$. The same form of solution holds for any of the three fields in the absence of any interactions. From this we see that the pulse duration (one-half the 1/e width of the amplitude function) after propagating a distance $z$ is given by $2T_p\sqrt{1+(\beta_{2p}z)^2/T_p^4}$, which at $z=0$ is $2T_p$. (The full-width at 1/e of the maximum values of the intensity function is $2.81\,T_p$.) This confirms, as mentioned earlier, that we can neglect the higher-order dispersion because we restrict our examples to satisfy $(\beta_{2j}z)^2/T_p^4 \ll 1$.

An important quantity is the conversion efficiency (CE) of signal mode to SFG mode, given by

$$\eta = \int_{-\infty}^{\infty} |A_r(L,t)|^2\,dt \ , \qquad (12)$$

which is bounded by 0 and 1, given that the input signal pulse is unit-normalized. For a single-photon input, $\eta$ is the probability that the photon is converted from the input mode to the SFG mode.

## 5. Temporal stretching and conversion efficiency

In this paper we illustrate SFG for various relationships among the group velocities of the three pulses for the process 1560 nm (signal) + 907 nm (pump) → 574 nm (SFG), for which SSGVM can be achieved by wave-guide engineering as pointed out in Fig.2, following [all17]. Thus, we take $\beta_p = 7.534$ ps/mm as for bulk PPNL and assume that $\beta_s = \beta_p$ unless stated otherwise. And we assume that the group velocity of the SFG field can be varied using slow-light techniques. In all cases the pump and signal pulses have duration parameter $T_p = T_2 = 1$ ps, so their full-width at half maximum of intensity equals around 3 ps, and we assume no chirp of the pump and input signal fields. We center these two input pulses at $T_{center} = 20$ ps.

To confirm quantitatively the points being made in Fig.1, we plot in **Fig. 3** the intensity (square of the solution in Eq.(4)) for $\beta_r = 4\beta_p = 30.14$ ps/mm for two values of coupling coefficient $\gamma$ and three values of medium length. In all the following, we state the values of $\gamma$ without restating its units ($s^{1/2}/m$). The temporal stretching and, at higher coupling strength temporal distortion, are observed as anticipated. The pulse shapes appear flipped compared to in **Fig. 1** because here we plot intensity versus time, not position. Also note the difference of the vertical axis scales, implying higher conversion efficiency for greater coupling coefficient. The small spike at early times results from the onset of oscillation in the Bessel function when its argument exceeds its first zero at 2.40583 and corresponds to a transient that occurs as the pump and signal pulses leave the output face of the crystal where the signal is nearly depleted. (The critical coupling strength for $\gamma = 1.4$ and $L = 40$ mm equals $\gamma_{crit} = 1.28$, from Eq.(9).)

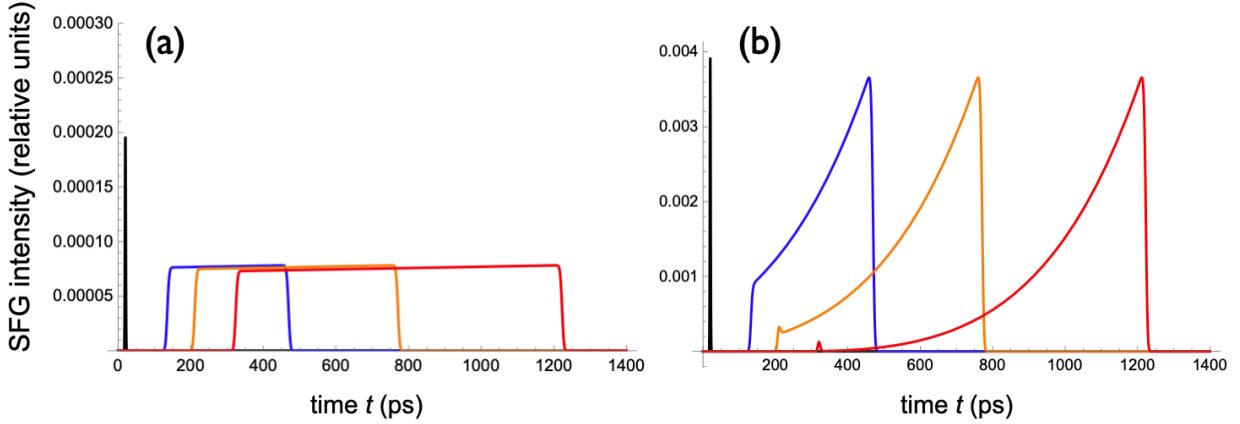

Fig. 3. (a) SFG intensity $|A_r(L,t)|^2$ versus time $t$ (ps) for $\beta_s = \beta_p = 7.534$ ps/mm, $\beta_r = 4\beta_r$, for $L$ = {15 mm (blue), 25 mm (orange), 40 mm (red)}, with coupling coefficient $\gamma = 0.2$, (b) Same but with coupling coefficient $\gamma = 1.4$. The not-to-scale input pump pulse (same as the signal pulse) is shown in black at $t$ = 20 ps.

To show why it is not a good idea to engineer all three group velocities to be equal, for **Fig. 4** we assume all the group velocities are essentially equal, with $\beta_p = \beta_s = 7.534$ ps/mm and $\beta_r = 1.001\beta_p$, where the factor of 1.001 is inserted to prevent a numerical divergence as a result of the $\bar{\gamma}$ prefactor in Eq.(4). We fix the medium length to $L$ = 25 mm and we see in Fig.4(a) that the solutions are oscillatory, as predicted by Eq.(10), with spatial oscillation period having different values in different time slices as a result of the temporal profile of the pump pulse. We also see oscillatory behavior of the conversion efficiency in Fig.(b), similar to that shown in [don16a]. More importantly there is no spectral compression, which would be indicated by a stretching in time. Instead, the oscillations tend to broaden the bandwidth. (recall we are assuming no pre-chirps of the pump and signal.)

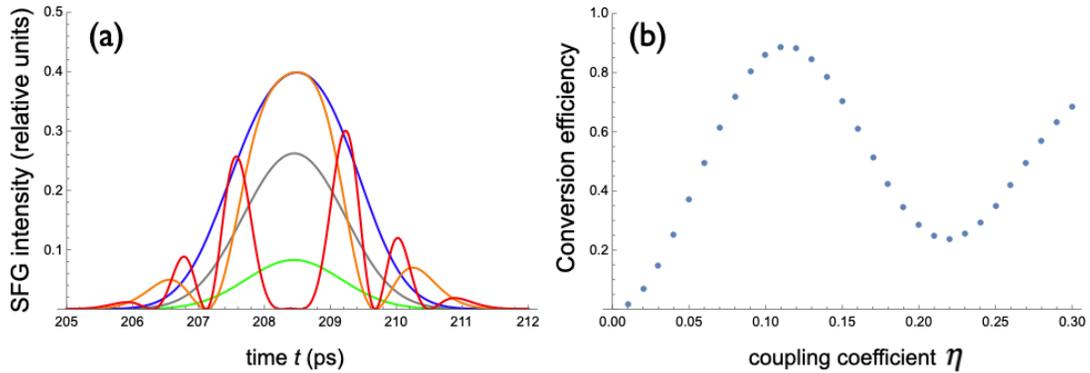

Fig. 4. (a) SFG intensity $|A_r(L,t)|^2$ versus time $t$ (ps) for $\beta_s = \beta_p = 7.534$ ps/mm, $\beta_r = 7.542$ ps/mm, $T_p = 1$ ps, $T_{center} = 20$ ps, $L$ = 25 mm, with varying coupling coefficient $\gamma$ = {0.03, green; 0.06, gray; 0.1, blue; 0.3, orange; 0.6, red}, (b) Conversion efficiency $\eta$ versus coupling coefficient $\gamma$ = 0 to 0.3.

In **Figs. 5** through **7** we illustrate the situation of principal interest, that is SSGVM, where the signal and pump are group-velocity matched, $\beta_s = \beta_p = 7.534$ ps/mm, while the SFG has slower and variable group velocity. Here we set the medium length to $L = 40$ mm.

In **Fig. 5** we assume the SFG has group velocity given by the realistic bulk PPLN value, $\beta_r = 8.132$ ps/mm $= 1.079\,\beta_p$. In this case we still see spatial-temporal oscillations, and moderate temporal stretching (spectral compression), but it degrades for higher CE. (In fact, these oscillations are what prevents the quantum pulse gate using bulk dispersion properties from reaching high conversion efficiency while maintaining good discrimination between orthogonal temporal modes. [red17])

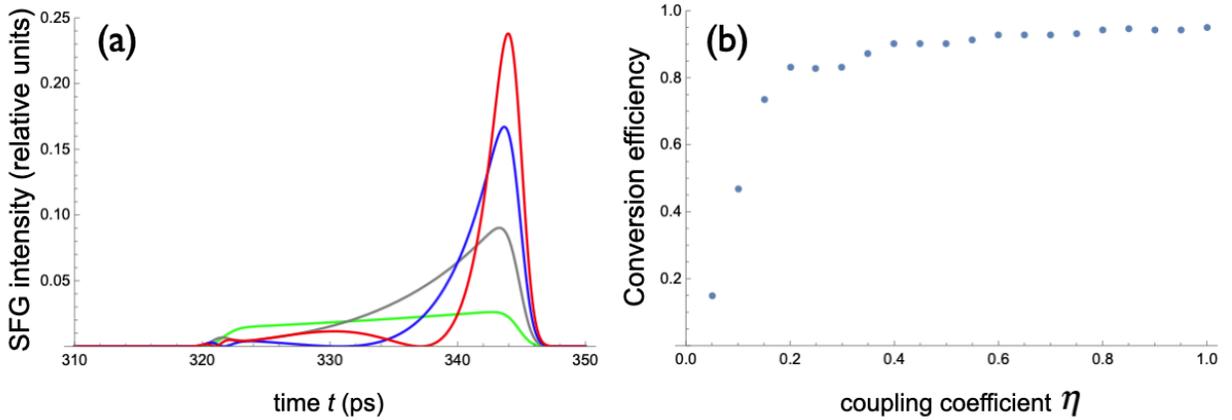

Fig. 5. 5 (a) SFG intensity $|A_r(L,t)|^2$ versus time $t$ (ps) for $\beta_s = \beta_p = 7.534$ ps/mm, $\beta_r = 8.132$ ps/mm, $T_p = 1\ ps$, $T_{center} = 20\ ps$, $L = 40$ mm, with varying coupling coefficient $\gamma = \{0.1,\ \text{green};\ 0.2,\ \text{gray};\ 0.1,\ \text{blue};\ 0.3,\ \text{orange};\ 0.4,\ \text{red}\}$, (b) Conversion efficiency $\eta$ for coupling coefficient $\gamma = 0$ to 1.

In **Fig. 6** we assume the SFG has group velocity corresponding to a slow-light wave guide with $\beta_r = 4\beta_p = 30.1$ ps/mm, while the signal and pump velocities are as before. In this case spatial-temporal oscillations are suppressed because the continual delay of the SFG field suppresses back-conversion from SFG to the signal field. The temporal duration is increased to the value $T_r = (\beta_r - \beta_s)L = 904$ ps (recall that this value is independent of the input signal duration, which we have taken in all cases to be $2.81\,T_p = 2.81$ ps for the full-width at 1/e of the intensity function. Here we have a temporal stretching factor $904/2.81 = 322$. The spectral compression factor is somewhat less than 322 as a result of the shape of the (chirp-free) pulses. The CE approaches 0.9 at the higher couple values, but the spectral bandwidth evidently increases at these higher CE values as a result of the relative shortening of the pulse duration. When the coupling parameter $\gamma$ exceeds the critical value $\gamma_{crit} = 1.28$ the Bessel function begins to oscillate, as seen the (red) curve for $\gamma = 2.0$, and the CE slightly increases.

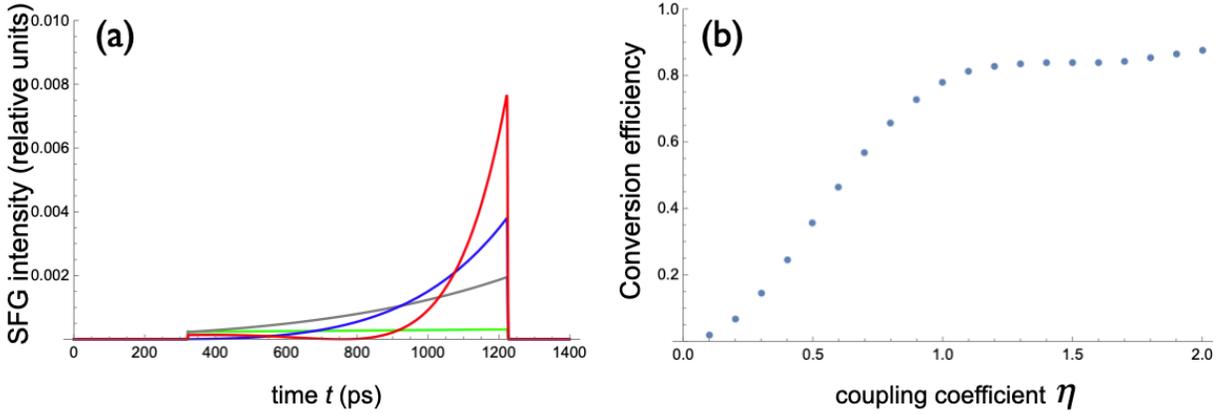

Fig. 6. (a) SFG intensity $|A_r(L,t)|^2$ versus time $t$ (ps) for $\beta_s = \beta_p = 7.534$ ps/mm, $\beta_r = 30.1$ ps/mm, $T_p = 1$ ps, $T_{center} = 20$ ps, $L = 40$ mm, with varying coupling coefficient $\gamma = $ {0.4, green; 1.0, gray; 1.4, blue; 1.949, red}, (b) Conversion efficiency $\eta$ for coupling coefficient $\gamma = 0$ to 2.

In **Fig. 7** we assume the SFG has group velocity corresponding to a slow-light wave guide with $\beta_r = 8\beta_p = 60.2$ ps/mm. We see that the spatial-temporal oscillations are further suppressed and the SFG evolves into a rising exponential-like shape, suitable for loading into an optical cavity. The temporal duration is increased to the value $(\beta_r - \beta_s)L = 2110$ ps. Here we have a temporal stretching by a factor 2110/2.81 = 751. Again, the spectral compression (inferred from the pulse shape) decreases at these higher CE values but is still significantly greater than can be achieved by other known schemes that use SFG in single-pass nonlinear optical media. Note the exponential-like shape for the largest coupling strength, which we set to $\gamma = 2$, above which the temporal mode begins to oscillate, as in **Figs. 5** and **6**. In the present figure we do not allow the coupling coefficient to exceed the critical value, $\gamma_{crit} = 1.949$, and we declare this value to correspond to the "optimal shape" of the SFG. Again, we note that increasing the coupling coefficient (e.g., through the pump intensity) further to increase the CE is not beneficial, as it impedes the spectral compression; in this sense the maximum useful CE is around 83% as a matter of principle.

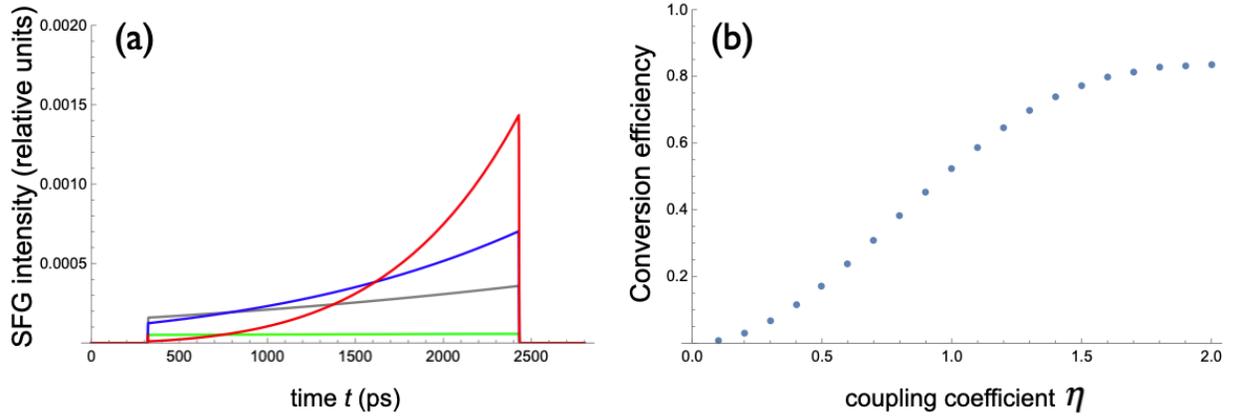

Fig. 7. (a) SFG intensity $|A_r(L,t)|^2$ versus time $t$ (ps) for $\beta_s = \beta_p = 7.534$ ps/mm, $\beta_r = 60.2$ ps/mm, $T_p = 1$ ps, $T_{center} = 20$ ps, $L = 40$ mm, with varying coupling coefficient $\gamma = \{0.4,$ green; $1.0,$ gray; $1.4,$ blue; $1.949,$ red$\}$, (b) Conversion efficiency $\eta$ for coupling coefficient varied from zero to $\gamma_{crit} = 1.94$.

Again, it should be noted that similar effects have been observed in second harmonic generation with strong fs pulses, where stretched, rising exponential-like shapes and accompanying spectral compression were observed. [mar07] In that case, with ultrashort pulses, slow-light engineering was not necessary, as bulk material dispersion is sufficient to create the needed group-velocity mismatch.

## 6. Spectral compression

For the lowest coupling strengths ($\gamma \ll \gamma_{crit}$), the SFG pulse is generated with a rectangular shape with duration $T_r$, and its spectrum is given by the square of its Fourier transform normalized to unity by area,

$$S_{rect}(\omega) = \frac{T_r}{2\pi} \frac{sin^2(\omega T_r / 2)}{(\omega T_r / 2)^2} \qquad (13)$$

Its full width (in units rad/s) can be defined as the frequency detuning from the peak at which the *sinc* function first goes to zero, that is $\delta\omega = 2\pi / T_r = 2\pi / (\beta_r - \beta_s)L$.

For moderate or high coupling strength (and thus high CE), no simple and accurate analytical form for the spectrum or its width seems to be known. In **Fig. 8** we show the numerically computed, unit-normalized energy (probability) spectrum corresponding to the case in Fig. 7 for $\gamma = 2$ (the exponential-like red curve in Fig. 7). And we plot the spectrum using Eq.(13) for the case in Fig. 7 for $\gamma = 0.4$ (the rectangular green curve in Fig. 7). We observe that the spectrum broadens when the CE increases to $\gamma = \gamma_{crit}$, but it does not exceed around twice the low-CE value (i.e., when $\gamma << \gamma_{crit}$).

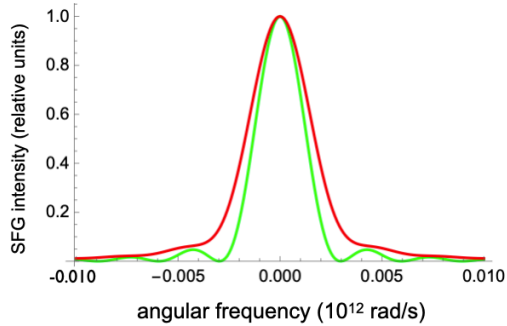

Fig. 8. Spectral intensity vs. angular frequency corresponding to two of the cases in Fig. 7. Green curve: $\gamma = 0.4$; Red curve: $\gamma = \gamma_{crit} = 1.949$. For $\gamma = 0.4$, the spectrum (green curve) is a sinc-squared function. For $\gamma = \gamma_{crit}$, the CE is around 83% and the pulse shape of the SFG is optimal (windowed rising exponential). In both cases, an estimate of the FWHM bandwidth is $\delta\omega \approx 2\pi / (\beta_r - \beta_s)L$, which in the present case equals 0.003.

## 7. Prospects for slow-light wave-guide engineering for spectral compression

The present proposal for improving the performance of mode conversion relies on the possibility to engineer the group velocities (GVs) of the pump, signal and SFG fields appropriately. Here the case is made that such is possible using current methods. Matching the GVs of the pump and signal fields is accomplished by designing the transverse profile of a wave guide (for example a ridge wave guide) so that modal dispersion becomes comparable to the bulk material dispersion, as in [all17] and references therein.

Engineering the GV of the SFG to be much slower than the pump and signal fields can be accomplished using design methods learned over many years of studying slow-light structures in waveguides with application in both linear and nonlinear optics. The general ideas of slow light are reviewed in [boy11]. Detailed reviews of the ideas of slow light in optical waveguides are presented in [bab08] and [sch10], where several crucial ideas are noted:
  1. Slow light (reduced GV) can be achieved in either photonic-wire waveguides (PWW) such as ridge waveguides or in planar photonic-crystal waveguides (PCW). In the former the slope of the dispersion curve $\omega(k)$ flattens slightly as the wavenumber ($k$) transitions from that in the waveguide cladding to that in the core. In the latter the slope of the dispersion curve approaches zero near the band edge of the photonic crystal.
  2. There is typically a trade-off between achievable pulse delay and the bandwidth over which a delayed pulse can be supported. The figure of merit for this consideration is the delay-bandwidth product (DBP). A large bandwidth is desired to avoid pulse distortion. In this regard it is desirable to engineer a wide spectral region of constant slope $d\omega / dk$ (that is, group velocity). Such a region is called a straight band or flat band and can be engineered in a PCW by careful selection of diameters and position of the air holes creating the photonic crystal bandgap.

    3. In principle, a waveguide can be engineered to have near-zero back reflection and side scattering, but in practice imperfections and material absorption will give rise to loss.

    4. Losses associated with input and output coupling can be minimized to less than -30 dB in theory by introducing the slow-light region adiabatically. Other methods exist as wel. [sch10]

    5. There exist design methods to minimize the effects of higher-order dispersion in slow-light wave guides.

A flexible and powerful platform for fabricating the needed slow-light SFG structure is thin-film lithium niobite (TFLN) having a ridge waveguide with corrugated side walls to create a series of coupled resonators. [che22] In that study a group index of 6.8 with slow-down factor of 3.0 relative to a simple ridge wave guide was achieved over a bandwidth of 4.3 nm. For a recent review of TFNL devices and their processing see [xin24]. In another study a group index of 30 was achieved in a PCW. [jav15]

## 8. Discussion

A slow-light scheme is proposed for simultaneous frequency conversion and temporal stretching and spectrum compression of a weak optical pulse, which may be in any quantum state including a single-photon state. In this general sense, the process implements mode conversion rather than state conversion. Such a process plays crucial roles in a number of schemes for constructing quantum networks. [ans18, aws21, sha24]

Assuming that appropriate slow-light waveguides can be fabricated, theoretical modeling shows that a 3-ps pulse can be converted into a pulse with duration in the ns regime with an intrinsic efficiency exceeding 80% (neglecting practical losses). The input pulse can be Gaussian or any smooth-shaped pulse and the converted pulse will have a near-exponential rising shape, which is suitable for temporal-mode matching into an optical cavity. Increasing the CE further is not beneficial, as it impedes the spectral compression; in this sense the maximum useful CE is around 83% as a matter of principle.

Note again that for large stretching, the duration of the stretched pulse is independent of the duration of the input pulse, assuming that pulse is not so short that higher-order dispersion would interfere with the process.

The present challenge is to design and fabricate a waveguide structure that both matches the pump and signal GVs while creating a much slower SFG field and minimizes higher-order dispersion. Given such a structure has been fabricated, its dispersion properties should be characterized experimentally to inform what spatial period the second-order nonlinearity should be polled with to achieve quasi-phase matching for the central frequencies.

**Funding**
This work was supported by the Engineering Research Centers Program of the National Science Foundation under Grant #1941583 to the NSF-ERC Center for Quantum Networks.

## Acknowledgments

Thanks to Marcus Allgaier, Benjamin Szamosfalvi, Ryan Camacho, C.J. Xin, Marko Loncar, Franco Wong, Jeffrey Shapiro, Brian Smith, and Clark Embleton for helpful discussions. This work was supported by the Engineering Research Centers Program of the National Science Foundation under Grant 1941583 to the NSF-ERC Center for Quantum Networks.

## Disclosures
The author declares no conflicts of interest.

## Data availability
Data underlying the results presented are fully contained in this paper. Calculations and plotting were performed using Mathematica.

## Appendix A Solution with full group-velocity matching

If the group-velocity differences and higher-dispersion can be neglected, we set $\beta_r = \beta_s = \beta_p = \beta$ and solve Eq.(1),

$$(\partial_z + \beta_r \partial_t) A_r(z,t) = i\gamma A_p(t - \beta z) A_s(z,t)$$
$$(\partial_z + \beta_s \partial_t) A_s(z,t) = i\gamma A_p^*(t - \beta z) A_r(z,t).$$

Changing variables to $\tau = t - \beta z$ and $x = z$, and defining $ie^{-i\theta(\tau)}\tilde{A}_r(x,\tau) \equiv \tilde{B}_r(x,\tau)$ gives

$$\partial_x \tilde{B}_r(x,\tau) = -\gamma |A_p(\tau)| \tilde{A}_s(x,\tau)$$
$$\partial_x \tilde{A}_s(x,\tau) = \gamma |A_p(\tau)| \tilde{B}_r(x,\tau),$$

with solution,

$$\tilde{B}_r(x,\tau) = \tilde{B}_r(0,\tau)\cos(\gamma|A_p(\tau)|x) - \tilde{A}_s(0,\tau)\sin(\gamma|A_p(\tau)|x)$$
$$\tilde{A}_s(x,\tau) = \tilde{A}_s(0,\tau)\cos(\gamma|A_p(\tau)|x) + \tilde{B}_s(0,\tau)\sin(\gamma|A_p(\tau)|x).$$

## References

[all17] M. Allgaier, V. Ansari, L. Sansoni, C. Eigner, V. Quiring, R. Ricken, G. Harder, B. Brecht, and C. Silberhorn, "Highly efficient frequency conversion with bandwidth compression of quantum light," Nat. Commun. 8, 14288 (2017).

[ans18] V. Ansari, J. M. Donohue, B. Brecht, and C. Silberhorn, "Tailoring nonlinear processes for quantum optics with pulsed temporal-mode encodings," Optica 5(5), 534–550 (2018).

[ave17] V. Averchenko, D. Sych, G. Schunk, U. Vogl, C. Marquardt, and G. Leuchs, "Temporal shaping of single photons enabled by entanglement," Phys. Rev. A 96(4), 043822 (2017).

[aws21] D. Awschalom, K. K. Berggren, H. Bernien, S. Bhave, L. D. Carr, P. Davids, S. E. Economou, et al., "Development of quantum interconnects (quics) for next-generation information technologies," PRX Quantum 2(1), 017002 (2021).

[bab08] T. Baba, "Slow light in photonic crystals," Nat. Photonics 2(8), 465–473 (2008).

[boy11] R. W. Boyd, "Material slow light and structural slow light: similarities and differences for nonlinear optics," JOSA B 28(12), A38–A44 (2011).

[bre14] B. Brecht, A. Eckstein, R. Ricken, V. Quiring, H. Suche, L. Sansoni, and C. Silberhorn, "Demonstration of coherent time-frequency Schmidt mode selection using dispersion-engineered frequency conversion," Phys. Rev. A 90, 030302 (2014).

[bre15] B. Brecht, D. V. Reddy, C. Silberhorn, and M. G. Raymer, "Photon temporal modes: a complete framework for quantum information science," Phys. Rev. X 5(4), 041017 (2015).

[che22] G. Chen and L. Liu, "Slow-light waveguide structure using coupled Bragg grating resonators on thin-film lithium niobate," in 2022 Asia Communications and Photonics Conference (ACP), IEEE, 1573–1575 (2022).

[che23] K. C. Chen, P. Dhara, M. Heuck, Y. Lee, W. Dai, S. Guha, and D. Englund, "Zero-added-loss entangled-photon multiplexing for ground- and space-based quantum networks," Phys. Rev. Appl. 19(5), 054029 (2023).

[cle16] S. Clemmen, A. Farsi, S. Ramelow, and A. L. Gaeta, "Ramsey interference with single photons," Phys. Rev. Lett. 117(22), 223601 (2016).

[don16a] J. M. Donohue, M. D. Mazurek, and K. J. Resch, "Theory of high-efficiency sum-frequency generation for single-photon waveform conversion," Phys. Rev. A 91(3), 033809 (2015).

[don16b] J. M. Donohue, M. Mastrovich, and K. J. Resch, "Spectrally engineering photonic entanglement with a time lens," Phys. Rev. Lett. 117(24), 243602 (2016).

[eck11] A. Eckstein, B. Brecht, and C. Silberhorn, "A quantum pulse gate based on spectrally engineered sum frequency generation," Opt. Express 19, 13770–13778 (2011).

[fab20] C. Fabre and N. Treps, "Modes and states in quantum optics," Rev. Mod. Phys. 92(3), 035005 (2020).

[jav15] A. Javadi, I. Söllner, M. Arcari, S. L. Hansen, L. Midolo, S. Mahmoodian, G. Kiršanskė, et al., "Single-photon non-linear optics with a quantum dot in a waveguide," Nat. Commun. 6, 8655 (2015).

[jun97] D. H. Jundt, "Temperature-dependent Sellmeier equation for the index of refraction, n_e, in congruent lithium niobate," Opt. Lett. 22(20), 1553–1555 (1997).

[kar17] M. Karpiński, M. Jachura, L. J. Wright, and B. J. Smith, "Bandwidth manipulation of quantum light by an electro-optic time lens," Nat. Photonics 11(1), 53–57 (2017).

[kum90] P. Kumar, "Quantum frequency conversion," Opt. Lett. 15(24), 1476–1478 (1990).

[lav13] J. Lavoie, J. M. Donohue, L. G. Wright, A. Fedrizzi, and K. J. Resch, "Spectral compression of single photons," Nat. Photonics 7(5), 363–366 (2013).

[li19] H. Li, H. Liu, and X. Chen, "Nonlinear frequency conversion of vectorial optical fields with a Mach-Zehnder interferometer," Appl. Phys. Lett. 114(24) (2019).

[man16] P. Manurkar, N. Jain, M. Silver, Y. P. Huang, C. Langrock, M. M. Fejer, P. Kumar, and G. S. Kanter, "Multidimensional mode-separable frequency conversion for high-speed quantum communication," Optica 3(12), 1300–1307 (2016).

[mar07] M. Marangoni, D. Brida, M. Quintavalle, G. Cirmi, F. M. Pigozzo, C. Manzoni, F. Baronio, A. D. Capobianco, and G. Cerullo, "Narrow-bandwidth picosecond pulses by spectral compression of femtosecond pulses in a second-order nonlinear crystal," Opt. Express 15(14), 8884–8891 (2007).

[mat16] N. Matsuda, "Deterministic reshaping of single-photon spectra using cross-phase modulation," Sci. Adv. 2(3), e1501223 (2016).

[mcg10] H. J. McGuinness, M. G. Raymer, C. J. McKinstrie, and S. Radic, "Quantum frequency translation of single-photon states in a photonic crystal fiber," Phys. Rev. Lett. 105(9), 093604 (2010).

[mck12] C. J. McKinstrie, L. Mejling, M. G. Raymer, and K. Rottwitt, "Quantum-state-preserving optical frequency conversion and pulse reshaping by four-wave mixing," Phys. Rev. A 85(5), 053829 (2012).

[pol24] R. Pollmann, F. Roeder, V. Quiring, R. Ricken, C. Eigner, B. Brecht, and C. Silberhorn, "Integrated, bright broadband, two-colour parametric down-conversion source," Opt. Express 32(14), 23945–23955 (2024).

[ray20] M. G. Raymer and I. A. Walmsley, "Temporal modes in quantum optics: then and now," Physica Scripta 95(6), 064002 (2020).

[ray23] M. G. Raymer and P. Polakos, "States, modes, fields, and photons in quantum optics," arXiv preprint arXiv:2306.07807 (2023).

[ray24] M. G. Raymer, C. Embleton, and J. H. Shapiro, "The Duan-Kimble cavity-atom quantum memory loading scheme revisited," Phys. Rev. Appl. 22, 044013 (2024).

[red13] D. V. Reddy, M. G. Raymer, C. J. McKinstrie, L. Mejling, and K. Rottwitt, "Temporal mode selectivity by frequency conversion in second-order nonlinear optical waveguides," Opt. Express 21(11), 13840–13863 (2013).

[red14] D. V. Reddy, M. G. Raymer, and C. J. McKinstrie, "Efficient sorting of quantum-optical wave packets by temporal-mode interferometry," Opt. Lett. 39(10), 2924–2927 (2014).

[red17] D. V. Reddy and M. G. Raymer, "Engineering temporal-mode-selective frequency conversion in nonlinear optical waveguides: from theory to experiment," Opt. Express 25(11), 12952–12966 (2017).

[red18] D. V. Reddy and M. G. Raymer, "High-selectivity quantum pulse gating of photonic temporal modes using all-optical Ramsey interferometry," Optica 5(4), 423–428 (2018).

[rik18] R. Ikuta, T. Kobayashi, T. Kawakami, S. Miki, M. Yabuno, T. Yamashita, H. Terai, et al., "Polarization insensitive frequency conversion for an atom-photon entanglement distribution via a telecom network," Nat. Commun. 9(1), 1997 (2018).

[sch10] S. A. Schulz, L. O'Faolain, D. M. Beggs, T. P. White, A. Melloni, and T. F. Krauss, "Dispersion engineered slow light in photonic crystals: a comparison," J. Opt. 12(10), 104004 (2010).

[sha24] J. H. Shapiro, M. G. Raymer, C. Embleton, F. N. C. Wong, and B. J. Smith, "Entanglement source and quantum memory analysis for zero added-loss multiplexing," Phys. Rev. Appl. 22, 044014 (2024).

[sko16] G. Skolianos, A. Arora, M. Bernier, and M. Digonnet, "Slow light in fiber Bragg gratings and its applications," J. Phys. D Appl. Phys. 49(46), 463001 (2016).

[sto09] M. Stobińska, G. Alber, and G. Leuchs, "Perfect excitation of a matter qubit by a single photon in free space," Europhys. Lett. 86(1), 14007 (2009).

[xin24] C. J. Xin, S. Lu, J. Yang, A. Shams-Ansari, B. Desiatov, L. S. Magalhães, S. S. Ghosh, et al., "Wavelength-accurate and wafer-scale process for nonlinear frequency mixers in thin-film lithium niobate," arXiv preprint arXiv:2404.12381 (2024).

[yan19] C. Yang, Z. Y. Zhou, Y. Li, Y. H. Li, S. L. Liu, S. K. Liu, Z. H. Xu, G. C. Guo, and B. S. Shi, "Nonlinear frequency conversion and manipulation of vector beams in a Sagnac loop," Opt. Lett. 44(2), 219–222 (2019).